*Probing the First Stars and Black Holes in the Early Universe with the Dark Ages Radio Explorer (DARE)*

Jack O. Burns[a,b], J. Lazio[c,b], S. Bale[d,b], J. Bowman[e,b], R. Bradley[f,b], C. Carilli[g,b], S. Furlanetto[h,b], G. Harker[a,b], A. Loeb[i,b], and J. Pritchard[j,b]

[a]*Center for Astrophysics & Space Astronomy, Department of Astrophysical & Planetary Sciences, 593 UCB, University of Colorado Boulder, Boulder, CO 80309, USA, jack.burns@colorado.edu*
[b]*NASA Lunar Science Institute, NASA Ames Research Center, Moffett Field, CA 94035, USA*
[c]*Jet Propulsion Laboratory, California Institute of Technology, M/S 138-308, 4800 Oak Grove Dr., Pasadena, CA, 91109, USA*
[d]*Space Sciences Laboratory, University of California at Berkeley, Berkeley, CA 94720 USA*
[e]*Arizona State University, Department of Physics, P.O. Box 871504, Tempe, AZ 85287-1504 USA*
[f]*National Radio Astronomy Observatory, 520 Edgement Road, Charlottesville, VA 22903 USA*
[g]*National Radio Astronomy Observatory, P.O. Box O, 1003 Lopezville Road, Socorro, NM 87801-0387 USA*
[h]*University of California at Los Angeles, Department of Physics and Astronomy, Los Angeles, CA 90095-1547 USA*
[i]*Center for Astrophysics, 60 Garden St., MS 51, Cambridge, MA 02138 USA*
[j]*Hubble Fellow, Center for Astrophysics, 60 Garden St., MS 51, Cambridge, MA 02138 USA*

**Abstract**

A concept for a new space-based cosmology mission called the Dark Ages Radio Explorer (DARE) is presented in this paper. DARE's science objectives include (1) When did the first stars form? (2) When did the first accreting black holes form? (3) When did Reionization begin? (4) What surprises does the end of the Dark Ages hold (e.g., Dark Matter decay)?  DARE will use the highly-redshifted hyperfine 21-cm transition from neutral hydrogen to track the formation of the first luminous objects by their impact on the intergalactic medium during the end of the Dark Ages and during Cosmic Dawn (redshifts z=11–35).  It will measure the sky-averaged spin temperature of neutral hydrogen at the unexplored epoch 80-420 million years after the Big Bang, providing the first evidence of the earliest stars and galaxies to illuminate the cosmos and testing our models of galaxy formation.   DARE's approach is to measure the expected spectral features in the sky-averaged, redshifted 21-cm signal over a radio bandpass of 40-120 MHz.   DARE orbits the Moon for a mission lifetime of 3 years and takes data above the lunar farside, the only location in the inner solar system proven to be free of human-generated radio frequency interference and any significant ionosphere.  The science instrument is composed of a three-element radiometer, including electrically-short, tapered, bi-conical dipole antennas, a receiver, and a digital spectrometer.  The smooth frequency response of the antennas and the differential spectral calibration approach using a Markov Chain Monte Carlo technique will be applied to detect the weak cosmic 21-cm signal in the presence of the intense solar system and Galactic foreground emissions.

## 1. Introduction

"What were the first objects to light up the Universe, and when did they do it?" These are amongst the most fundamental questions in modern astrophysics and cosmology as stated recently by the Astro2010 Decadal report, *New Worlds, New Horizons in Astronomy and Astrophysics* (NRC, 2010). The birth of the first stars and black holes—the end of the *Dark Ages* or *Cosmic Dawn*—is a transformative event in the history of the Universe. This epoch provides the key connection between telescopic images of the structures and galaxies seen at redshifts z<5 and observations of the extraordinarily smooth Universe 400,000 years after the Big Bang (z~1100) seen via the Cosmic Microwave Background (e.g., Samtleben et al. 2007). In this paper, we describe a mission concept called the Dark Ages Radio Explorer (DARE) that will investigate this early epoch of the Universe (~80–420 million years after the Big Bang) which was heretofore unobservable. DARE will produce the first sky-averaged, low frequency radio (40-120 MHz) spectrum arising from the highly redshifted 21-cm signal of neutral hydrogen that uniquely measures the times when the first stars and black holes appeared in the Universe.

These observations are challenging because the expected 21-cm signal strength is considerably below that of several foreground signals. However, DARE eliminates the most intense foreground, namely human-generated radio frequency interference (RFI), by being in a lunar orbit that takes it into the shielded zone above the lunar farside. Science data are acquired only when DARE is above the lunar farside, which is the only location in the inner solar system proven to be free of RFI by Radio Astronomy Explorer-2 (RAE-2, Alexander & Kaiser 1976).

This concept draws on a rich science and technology heritage from a ground-based pathfinder, the Experiment to Detect the Global Epoch of Reionization Signal (EDGES, Bowman & Rogers 2010a). Further, there is a long history of space-based RF systems operating in or near the DARE frequency range, e.g., the Earth remote sensing satellites FORTE, the GeoSAT Follow-On (GFO), and the lunar-orbiting RAE-2 (Alexander & Kaiser 1976).

In this paper, the theoretical drivers, a description of the Science Instrument, and the calibration of the radiometer in the presence of intense foregrounds are presented. The theoretical predictions of the sky-averaged spectrum corresponding to the epoch of the first stars, the first black holes, and the beginning of reionization are described in Section 2. In Section 3, we discuss why observations above the radio-quiet lunar farside are ideal to achieve the first detection of the weak, redshifted 21-cm signals. In Section 4, the DARE radiometer system consisting of tapered biconical dipole antennas, a low frequency receiver, and a digital spectrometer is outlined. In Section 5, the calibration and removal of intense cosmic foregrounds is described using differential spectral radiometry. In Section 6, constraints on the epochs of first star and black hole formation from the proposed DARE observations are presented. In Section 7, we conclude with a summary of the DARE mission concept.

## 2. Theoretical Motivation

*2.1 Tracing the First Luminous Objects with the Sky-Averaged 21-cm Spectrum*

The Big Bang produced a hot, dense, nearly homogeneous Universe. As the Universe expanded and cooled, particles, then nuclei, and finally atoms formed. At a redshift of about 1100, equivalent to about 400,000 years after the Big Bang, when the primordial plasma filling the Universe cooled sufficiently for protons and electrons to combine into neutral hydrogen atoms, the Universe became optically thin whereby photons from this early era no longer interacted with matter. We detect these photons today as the cosmic microwave background (CMB). The CMB shows that the Universe was still remarkably smooth and uniform, with fluctuations in the mass density of only a few parts in $10^5$ (Hu 2008). The Cosmic Background Explorer (COBE), the Wilkinson Microwave Anisotropy Probe (WMAP), and a number of ground-based telescopes have mapped this era with extraordinary precision, providing a detailed understanding of this early phase of the Universe's history (e.g., Dunkley et al. 2009).

After the protons and electrons combined to produce the first hydrogen atoms, the Universe consisted of a nearly uniform, almost completely neutral, intergalactic medium (IGM) for which the dominant matter component was hydrogen gas. With no luminous sources present, this era is known as the *Dark Ages*. Theoretical models predict that, over the next few hundred million years, gravity slowly condensed the gas into denser and denser regions, within which the first stars eventually appeared, marking *Cosmic Dawn*, at a redshift $z \approx 30$ (Bromm & Larson 2004). As more stars formed, and the first galaxies assembled, they flooded the Universe with ultraviolet photons capable of ionizing hydrogen gas. A few hundred million years after *Cosmic Dawn*, the first stars produced enough ultraviolet photons to ionize essentially all the Universe's hydrogen atoms (Furlanetto et al. 2006, and references therein). This *Reionization* era is the hallmark event of this early generation of galaxies, marking the phase transition of the IGM back to a nearly completely ionized state, which is now known to be well underway and potentially nearing completion at a redshift $z \approx 6$ (Fan, Carilli, & Keating 2006).

The beginning of structural complexity in the Universe constituted a remarkable transformation, but one that we have only just started to investigate observationally. Existing instruments (e.g., *Hubble Space Telescope* or HST) show a host of galaxies filling the Universe from about 500 million years after the Big Bang to the present day (i.e., *z*<10, Bouwens et al. 2010). Interestingly, although these galaxies appear many times smaller than those at lower redshifts, their stellar populations are fundamentally the same, though enriched to a lower level of heavy elements (e.g., McLure et al. 2011). By pushing even farther back, the truly first structures in the Universe can be studied. Theoretical models suggest that existing measurements (from *HST* and WMAP) are beginning to probe the tail end of Reionization, and the James Webb Space Telescope (JWST) will observe some of the earliest galaxies, possibly out to *z*~15 (Windhorst et al. 2006), while the Atacama Large Millimeter Array (ALMA) will trace the molecular content of galaxies possibly to *z*~10 (Carilli et al. 2008). Even so, the first stars and galaxies – in the *Dark Ages* and the *Cosmic Dawn* – currently lie beyond our reach.

The measurement approach adopted by DARE is to track the influence of the first stars and first accreting black holes on the neutral IGM by means of the intensity of the hydrogen hyperfine (or "spin-flip") transition. This transition originates from magnetic moments or spins of the proton and electron—if the angular momenta are aligned, the atom lies at a slightly higher energy than if they are anti-aligned (Furlanetto et al. 2006). The energy difference corresponds to a photon of rest wavelength of 21 cm or a rest frequency of 1420 MHz. Because of the expansion of the Universe, these photons from earlier epochs in the Universe are "redshifted" ($z$) to lower frequencies, $\nu_{obs}$=1420/(1+$z$) MHz (e.g., at $z$ = 30, $\nu$= 45 MHz). Importantly, the redshifting of the photons allows us to examine the history of the Universe as a function of time (each frequency corresponds to a different redshift or epoch of the Universe). Consequently, observations of the spin-flip line allow the *Dark Ages* and *Cosmic Dawn*—when the IGM was mostly neutral hydrogen—to be probed. Moreover, the 21-cm signal complements probes of other telescopes (e.g., *HST*, *JWST*, ALMA), as it provides an *indirect* measure of the effects of distant galaxies on the IGM. This allows it to measure the properties of stars and black holes inside galaxies that are much too small to be imaged directly.

Ground-based low-frequency radio telescopes operating at $\nu$>100 MHz, such as the Experiment to Detect the Global Epoch of Reionization Signal (EDGES, Bowman & Rogers 2010), the Murchison Widefield Array (MWA, Lonsdale et al. 2009), the Precision Array to Probe the Epoch of Reionization (PAPER, Parsons et al. 2010), and the Low Frequency Array (LOFAR, Harker et al. 2010) will study the end of Reionization. Table 1 summarizes on-going ground- and space-based work and compares these to DARE. Clearly, DARE will probe the redshifted 21-cm signal of hydrogen at higher redshifts ($z$ < 35), where the earliest signals produced by the first stars and accreting black holes, as well as the first stages of reionization, are expected to occur.

Throughout this paper, we use the radio astronomy convention relating power measured by a detector $P$ to a brightness temperature $T_B$,

$$P = k_B T_B \Delta\nu \quad (1)$$

where $k_B$ is Boltzmann's constant and $\Delta\nu$ is the frequency bandwidth over which the measurement is made. The 21 cm background signal brightness temperature is (Madau et al. 1997; Furlanetto et al. 2006),

$$T_b = 25 x_{HI} \left(\frac{1+z}{10}\right)^{1/2} \left(1 - \frac{T_{CMB}}{T_S}\right) \text{mK}, \quad (2)$$

Here, $x_{HI}$ is the fraction of hydrogen that is neutral (because spin-flip radiation only comes from neutral gas), $z$ is the redshift being probed, $T_{CMB}$ is the temperature of the CMB photon blackbody, and $T_S$ is the "spin temperature" of the gas. The spin temperature is a measure of the excitation state of the gas, providing the ratio of level populations for the ground and excited states in this transition (with the ground state more highly populated if the spin temperature is small). Its amplitude, and hence the strength of the 21-cm spectral line, depends on the balance between three competing processes: atomic collisions, absorption of CMB photons (and stimulated emission from that photon field), and the ultraviolet radiation background, which in turn reflect processes related to the origin of stars and accreting black holes.

Over the redshift range of interest, this last factor contains most of the physical information that DARE will measure. Figure 1 shows a model for the sky-averaged, redshifted 21-cm spectrum during the *Dark Ages,* with the brightness temperature measured relative to the CMB (Pritchard & Loeb 2010, see also Furlanetto 2006). The brightness temperature evolves with redshift, with several key **Turning Points** (i.e., inflection points) predicted in the spectrum. These Turning Points mark important physical events (Figure 1), which are due to the interplay between the first luminous sources and the physics of the hydrogen gas. In the following sections, we summarize the various physical effects that produce the Turning Points, following the approach of Pritchard & Loeb (2010). We stress that the physical processes themselves are based on well understood physics (e.g., quantum mechanics), but their timing and the rate at which they occur (or the resulting strength of the 21-cm signal) are quite uncertain. Thus, observations are crucial to constrain these important events during this epoch.

*2.2 DARE's Key Science Questions*

The key physical quantities that DARE can measure in order to test this hypothesis are described below in the context of the simple reference model presented in Figure 1, which is a "best guess" for the physics of these early galaxies given current information. We will also describe the wide range of plausible variations still allowed by theoretical models and how they influence the signal. We stress again that our summary reflects the current best understanding of this era, but this understanding is derived entirely from theoretical models. DARE's objective is to provide an observational basis for these discussions.

*2.2.1 When did the first stars ignite?*

Before the first stars and galaxies formed, the hydrogen gas was influenced only by atomic collisions, absorption of CMB photons, and gravitational effects. The gas cooled rapidly as the Universe expanded, to the point that the spin temperatures were *lower* than that of the CMB. At sufficiently early times, the high gas density caused a large rate of atomic collisions, so that the spin temperature approached the gas temperature, causing the 21-cm signal to appear in absorption against the CMB. However, as the Universe expanded, the decreased gas density reduced the collision rate, and interactions with CMB photons drove the spin temperature back into equilibrium with the CMB. This caused the spin-flip absorption signal to decrease and eventually disappear by $z \sim 30$ (between Turning Points A and B), because there is no longer any relative absorption or emission against the CMB.

Current predictions are that, shortly thereafter, the first stars appeared ($z \sim 30$, Turning Point B). Their ultraviolet photons scattered throughout the intergalactic medium and "turned on" the spin-flip 21-cm signal through the so-called Wouthuysen-Field mechanism (Wouthuysen 1952; Field 1959). Photons that redshift into the Lyman-α transition of neutral hydrogen cause the electron to jump from the ground state to the first excited level, from which they rapidly decay back to the ground state. However, the quantum mechanical selection rules allow the electron to occupy a different hyperfine state after this excitation and decay than before. When enough such scatterings occur, these interactions determine the spin temperature and bring it back into equilibrium with the gas temperature (still much colder than the CMB). Thus, the first stars produced a deep absorption trough (between Turning Points B and C, e.g., Furlanetto et al. 2006).

By determining the redshift of this transition, DARE will make the first measurement of the onset of star formation in the Universe. Interestingly, theoretical models predict that these stars form very differently than those near Earth: without heavy elements, the chemistry of this primordial gas differs from that in later generations of galaxies, likely causing the first stars to be tens or even hundreds of times larger than nearby stars (Bromm & Larson 2004). Observational probes of these first luminous objects are essential for understanding these first phases of structure formation.

Equally interesting is the transition from these exotic stars to more normal stars, which depends on several poorly-understood feedback mechanisms, like chemical enrichment, supernova feedback, and the ultraviolet radiation field (e.g., Furlanetto et al. 2006). This transition determines the slope at which the absorption deepens, between Turning Points B and C (Fig. 1), and DARE's measurement will provide a robust measurement of the radiative feedback in particular (Holzbauer & Furlanetto 2011).

Figure 2 illustrates the sensitivity of the spin-flip 21-cm signal to the ultraviolet radiation of the first stars. The solid black curve is identical to the model in Figure 1, but the blue curves vary the ultraviolet luminosity of the first stars by a factor of 100. Measuring the redshift of Turning Point B, and the shape of the subsequent absorption trough between Turning Points B and C, therefore constrains the formation mechanism of these stars and their properties.

*2.2.2 When did the first accreting black holes form?*

Black holes likely formed shortly after the first stars. Their intense gravity would have strongly accelerated any gas falling onto them, shock heating it to high temperatures and inducing strong X-ray emission. In contrast to ultraviolet photons, which have an extremely small mean free path into neutral hydrogen gas, X-ray photons can penetrate to great depths, but they eventually scatter through the gas and deposit their energy in a mixture of ionization, heat, and collisional excitations. This X-ray background very quickly heated the hydrogen gas well above its initial temperature (which was only $T \sim 10$ K). This heating transformed the spin-flip signal from absorption into emission as the gas became hotter than the CMB (between Turning Points C and D). Thus, by measuring the onset of gas heating at Turning Point C, DARE detects the earliest evidence of black hole accretion in the Universe.

The nature of the first black holes remains even more of a mystery than the first stars, forming either as ~10-100 solar mass remnants from those stars' explosions (e.g., Volonteri et al. 2008) or in much larger seeds (up to ~10,000 solar masses) via the direct collapse of gas in unusual environments where star formation was ineffective (Bromm & Loeb 2003). In either case, they must have accreted material very quickly in order to produce the ultra-luminous "quasars," powered by accretion into black holes a billion times more massive than the Sun, that are visible as early as 1 billion years after the Big Bang (Fan et al. 2006). The growth of these black holes determines the slope of the DARE spectrum between Turning Points C and D, during which the signal passes from absorption to emission. DARE thus offers an unprecedented opportunity to constrain the physics of black hole formation during the Cosmic Dawn.

The red curves in Figure 2 show how the spin-flip signal depends on X-ray heating. Only stellar remnant black holes are considered, and their X-ray luminosity is varied by a factor of 100 on either side of the reference model (which is calibrated to the X-ray emissivity of nearby galaxies; Pritchard & Loeb 2010).

*2.2.3 When did reionization begin?*

Once the gas became hot, the emission saturated, becoming independent of the precise spin temperature (equation [1]). The spin-flip signal remains roughly constant until photons from these stars and black holes started ionizing the gas in giant bubbles within the IGM. The rapid destruction of the neutral gas during *reionization* then eroded the spin-flip signal (between Turning Points D and E). DARE will constrain the onset of this reionization process and hence the first era of rapid star formation.

The timing of this stage depends on the overall ionizing efficiency of galaxies, which in turn depends on a large number of uncertain physical factors (for example, the rates at which galaxies form and cycle gas into stars, as well as the masses of these stars, and the rates at which these photons escape their host galaxies; see, e.g., Furlanetto et al. 2006). Existing measurements from the WMAP satellite suggest that the midpoint of reionization occurred at $z \sim 10$ (Dunkley et al. 2009). The slope of the DARE spectrum beyond Turning Point D tells us how these factors evolved, on average, through the early phases of galaxy formation. In combination with detailed measurements of individual objects from telescopes like JWST and ALMA, DARE will answer fundamental questions about the buildup of galaxies during the *Cosmic Dawn*.

*2.2.4 What surprises does the end of the Dark Ages hold?*

The standard cosmological paradigm that we have just described paints a "broad brush" picture of what is expected in the "missing" billion years between the formation of the CMB and the most distant galaxies currently visible. While observations of the CMB and galaxies some one billion years after the Big Bang strongly suggest that DARE will see the first stars and black holes, the unique information that DARE provides may point us toward surprising new insights into the earliest phases of structure formation in the Universe.

Until such observations become possible, this time period is truly the new "universum incognitum," the last blank space left on humankind's map of the Universe. DARE provides a key test of the physics of this era, possibly challenging our most basic cosmological assumptions. Did neutral hydrogen fill the Universe at this time? Did radiation sources similar to those that we expect form during the era, $z \sim 11$-$35$? Did they transform the hydrogen gas between these galaxies by illuminating, heating, and eventually ionizing it? Or will unexpected discoveries completely upend this picture?

Such surprises are impossible to predict, but the spin-flip signal is so sensitive to radiation backgrounds in the early Universe that DARE could reveal important deviations from our basic cosmological model. Figure 3 illustrates how the signal is affected in some such alternate scenarios. For illustrative purposes, we consider several different scenarios here (see also, e.g., Furlanetto, Oh, & Pierpaoli 2006, Mack & Wesley 2008, Natarajan & Schwarz 2009). The red line shows the signal if an unknown radiation background heats the intergalactic gas to high temperatures and couples the spin temperature to it. The dashed line shows a model in which dark matter annihilates over a long time scale, heating the gas significantly. The blue line shows a model without an ultraviolet background, rendering the Wouthuysen-Field mechanism ineffective and the spin-flip line nearly invisible. Finally, the dotted line turns off X-ray heating, allowing the IGM to remain cold and visible in strong absorption.

While all of these examples seem implausible today, they illustrate the power of exploring an entirely new era within the Dark Ages. By observing in this previously unexplored redshift range, DARE can confirm current models of the first luminous objects or measure deviations from models that might illuminate new physics.

## 3. Why DARE Needs to Orbit the Moon

A significant hurdle to a terrestrial global 21-cm experiment is human-generated RFI. Our civilization emits the vast majority of its integrated radio power at DARE frequencies (Loeb & Zaldarriaga 2007). At the Murchison Radio Observatory in Western Australia, a government-protected site, several FM radio signals are persistently detected by EDGES at the 1 K level (1000 times the limit needed to detect the Turning Points in Figure 1) due to reflections from meteors and aircraft (Bowman et al. 2008; Rogers & Bowman 2008; Bowman & Rogers 2010a,b; EDGES memo series[1]). Reflections from density structures in the ionosphere are also possible (Lazio et al. 2010).

In low Earth orbit (LEO), the effective antenna temperature from a single terrestrial FM transmitter in direct line-of-sight to the spacecraft could exceed $10^9$ K in a 1 MHz spectral channel, as compared to the 1 mK limit that we require for science observations (see Section 4). In geosynchronous Earth orbit (GEO), the effective antenna temperature would be reduced to only ~$10^8$ K. At the lunar L2 Lagrange point, ~60,000 km behind the Moon, the lunar disk does not sufficiently occult the Earth, because of diffraction of radio waves around the lunar limb. A distance of at least 4 AU is required to prevent DARE from detecting terrestrial RFI above 1 mK.

An alternate approach would be to equip the DARE spacecraft with a metallic screen designed to shield the spacecraft, and the science instrument, from terrestrial RFI. This approach would be analogous to the Sun shield planned for JWST. While such an RFI shield is in principle possible, e.g. achieving 50 dB of shielding, diffraction about the shield would still result in an effective antenna temperature of 4000 K if DARE were in LEO and 200 K at the lunar L2 point. Further, such a shield would be massive (increasing launch costs) and almost certainly require deployment, thereby increasing mission risk.

Lunar orbit is a far more practical solution. Further, both RAE-2 and the Apollo Command Modules, which had RF systems that bracket DARE frequencies, observed complete cessation of terrestrial RF emission when they passed into the radio-quiet zone above the lunar farside. Finally, reflections of terrestrial RFI from other spacecraft in view of DARE while in the radio quiet zone will be at a negligible level (< 1 nK). Placing DARE above the farside of the Moon removes all complications from Earth's ionosphere and terrestrial RFI, and blocks heliospheric emissions from the Sun for part of the orbit.

---

[1] http://www.haystack.mit.edu/ast/arrays/Edges/EDGES_memos/EdgesMemo.html.

DARE's position behind the Moon opens the entire RF spectrum to astronomical use, simplifies the receiver design, improves its inherent performance, and enables the science objectives. Completely eliminating RFI (from the Earth and the spacecraft environment) is the only guarantee that the redshifted 21-cm signal will not be corrupted by uncharacterized spectral artifacts. By lowering the spectral dynamic range from >$10^9$ to the sky limit of $10^6$, DARE avoids the bit-depth, linearity, clock stability, and digital noise challenges experienced by ground-based experiments.

DARE provides an ideal environment for the science antenna. The potential for far-field scattered radiation and uncertainties in the operational conditions are greatly reduced, allowing for the antenna impedance and primary beam pattern to be more stable, accurately modeled, effectively calibrated, and better matched to science goals. Control and characterization of antenna properties are the most important factors for subtracting astronomical foregrounds (Sections 4.4 and 5).

Finally, the motivation for DARE parallels the successful history of cosmic microwave background (CMB) anisotropy experiments. CMB studies are accomplished from the ground or balloons, but the space environment substantially improves scientific return. Terrestrial CMB instruments including BOOMERANG, DASI, CBI, and others developed fundamental techniques and illuminated key scientific uncertainties, but space missions COBE, WMAP, and Planck have capitalized on the space environment to provide definitive measurements. DARE builds on ground-based heritage, uses the ideal properties above the lunar farside to eliminate RFI, and provides a well characterized science antenna in order to achieve the best possible measurement of the redshifted 21-cm global signal.

## 4. The DARE Radiometer

The DARE science objectives require a carefully calibrated, load switching radiometer to measure radio frequency (RF) spectra over the 40–120 MHz range (Figure 4) of the VHF band. This radiometer consists of a dual-polarized antenna with a compact, integrated, front-end electronics package, a single-band, dual-channel receiver, and a digital spectrometer. The heritage of VHF band radiometry spans a half-decade to include ground-based experiments such as EDGES and PAPER, Earth remote sensing satellites like FORTE and GeoSAT Follow-On (GFO), and lunar orbiting spacecraft such as the Radio Astronomy Explorer-2 (Explorer 49). While the specific design of the DARE instrument is new, all subsystem components have a strong heritage with a technology readiness level (TRL) of at least 6 (TRL ≥ 6) on the standard scale.

To extract the desired science results at the significance levels described in Section 2, the radiometer must be well calibrated (Section 4.4). The DARE instrument design is guided by years of experience in calibrating spaceflight radiometers such as AMR and ground-based experiments such as EDGES and PAPER. Challenges addressed by the design include adequately characterizing the antenna power pattern, understanding the receiver noise contributions, and maintaining sufficient linearity to extract the 21-cm spectral features in the presence of dominant foregrounds.

Sensitive radiometers require that all supporting subsystems, including everything from Command and Data Handling (CDH) to the power supplies, must be RFI-quiet so as not to adversely affect the science measurements. However, two additional requirements are imposed on the DARE radiometer: 1) the system must be stable enough to permit long-duration, statistical measurements over many months of observing, and 2) the spectral response is smooth so that residual features upon calibration are small enough to permit unambiguous, noise-limited localization of the Turning Point frequencies. These special requirements together with the additional constraint that the science instrument will only operate when the spacecraft is above the lunar farside, makes DARE a rather novel radiometer system.

*4.1 Science Requirements*

The design of the DARE radiometer, from the RF transfer characteristics to the data management rates and operational features, is driven by the basic science requirements. In this section, we summarize only those RF characteristics that affect the science measurements. These include frequency range, spectral resolution, spectral response, and sensitivity.

**Frequency Range**: This requirement is set specifically by the goal of measuring the redshifts at which Turning Points B, C, and D occur. Current theoretical models suggest that these Turning Points should occur over the approximate redshift range of 11 to 35. Given the relation between rest-frame and observed frequency, this redshift range translates to a frequency range of 40 to 120 MHz, which is the design requirement for the Radiometer. We have not attempted to measure the redshift of Turning Point E because there are ground-based instruments that are likely to constrain the properties of the intergalactic medium at the end of the Epoch of Reionization before or on the same time scale as DARE becomes operational. Early in the concept development phase, we decided not to attempt to measure the redshift of Turning Point A because the mission duration would be excessively long. We discuss foregrounds and sensitivity below, but, briefly, at lower frequencies, foregrounds become stronger, necessitating ever longer mission durations.

**Spectral Resolution**: This requirement is set by the characteristic scale over which the Turning Points occur. Current theoretical models suggest that a useful measure would be 1 MHz. However, recombination lines from carbon atoms have been detected within or just below the DARE frequency range. These recombination lines are much narrower than 1 MHz. Estimates of their amplitude are uncertain as they depend upon the poorly known distribution of recombination line hosting gas within the Galaxy. Nonetheless, it is possible that recombination lines could produce a signal comparable to that expected for the Turning Points (Peters et al. 2011). In order to have sufficient spectral resolution to identify, and possibly mask recombination lines during the analysis, we specify a requirement of 10 kHz.

**Spectral Response**: Because the Universe is assumed to be homogeneous and isotropic, the sky-averaged 21-cm spectrum that DARE seeks to measure should vary only in frequency, but not with position on the sky. Conversely, the (stronger) foregrounds are expected to vary with position on the sky, but not spectrally. Consequently, we impose the requirement that the spectral response of the instrument be sufficiently smooth so that it does not vitiate the identification of the 21-cm spectrum.

**Sensitivity**: This requirement is set by the expected amplitudes of the Turning Points and the need to have an adequate signal-to-noise ratio. As can be seen from Figure 1, the amplitudes, relative to the CMB, vary for the different Turning Points. We adopt a value of 1 mK, in a 1 MHz band, as the sensitivity requirement, with this value being constant across the DARE frequency range. This value should provide a high signal-to-noise ratio detection for Turning Points C and D and should be adequate to identify the location of Turning Point B (see Section 5). Indeed, even if Turning Point B lies outside of the DARE frequency range, with this sensitivity requirement, we would still be able place some constraints on Turning Point B's location by using the shape of the measured spectrum.

*4.2 Antenna*

The Science Antenna will be a pair of bi-conical dipoles (Figure 5), supported at its periphery by a low-mass dielectric structure and at the feed points by a cylindrical mast. The antenna is made unidirectional to a certain degree by a set of deployable radials attached to the spacecraft bus that act as an effective ground plane. The orthogonally arranged dipoles will increase sensitivity by receiving both polarizations, provide an independent check on system performance, and enhance reliability through redundancy. Initial electromagnetic simulations and verification indicate that the antenna impedance and spatial power pattern are stable over time and vary smoothly as functions of frequency. The antenna design summarized below can be constructed from standard spacecraft materials and heritage components.

The key trades for the antenna design were among mass, volume, and electrical performance. A bi-conical dipole is a compact and low mass structure, yet also achieves the electromagnetic performance specified by the science requirements (Section 4.1). Alternate topologies such as high gain, aperture-type configurations and log-periodic antennas were excluded on the basis of their large size, the strong dependency of their terminal impedance on frequency, or both.

An electrically short dipole ($< 0.6\lambda$) will be used, with the half-wavelength resonance designed to occur near the upper end of the band. Consequently, the power pattern remains symmetrical, and both it and the terminal impedance should vary smoothly with frequency; the overall gain is lower than that of a resonant dipole but is sufficient given the strong Galactic emission at these frequencies. The combination of large diameter and tapering (1) lowers the dipole's Q (quality factor) to increase its bandwidth; (2) reduces reactive effects due to the series inductance at the end of the arm or arm-to-arm capacitance across the feed terminals; and (3) pushes any resonances due to circuit parasitic effects across the antenna terminals above the 40–120 MHz range.

The dipole cones will be constructed from an internal wire frame covered with standard gold-plated wire mesh used on deployable communication antennas. The dielectric support will be an Astroquartz/Kevlar fiber and cyanate ester resin composite, which has good thermal stability, low

RF loss, and has heritage from Deep Impact and the Mars Exploration Rovers (MER) (Figure 5). Further, this dielectric support will have a high resistivity, RF transparent Ge coating in order to bleed off any solar wind induced charge.

To improve the antennas' electrical performance, an effective ground plane will be formed by deployable radials extending outward from the spacecraft body and located approximately a quarter-wavelength (at the resonant frequency) from the dipole plane. The radials increase the forward gain, thereby (1) limiting the spatial extent of the sky model needed for calibration; (2) reducing the effect of asymmetrical spacecraft structures; and (3) providing modulation in detected power with spacecraft orientation. Radials are used to eliminate the need for a large monolithic structure, which would pose a significant deployment challenge. By contrast, small-diameter metallic radials can function as an effective ground plane, providing about 15 dB front-to-back ratio for 24 radials and have been deployed in hundreds of space missions without a single failure.

Modeling of the antenna indicates that it can be characterized to the level required to meet the science objectives and as specified by the science requirements. The electromagnetic performance of the antenna, radials, and a hexagonal spacecraft were modeled using CST Microwave Studio. Each dipole arm is 59 cm in length, with a taper of 40° to provide a smooth instrument response, and an 8.3° rotation away from the spacecraft to accommodate the likely launch vehicle fairing. The terminal ends are 65 cm from the 24 ground radials, each of which extends 2 m from the spacecraft. The model also included (1) a cylindrical mast and a metal conduit running from the spacecraft to the dipole feed points; (2) non-conductive support structures near the ends of the arms; (3) the front-end electronics; and (4) the propellant tank at the ground plane.

The power pattern is a single-lobed, smoothly varying spatial function (Figure 6), and the minimum directivity occurs off the ends of the dipole arms. The terminal impedance follows a well-behaved function of frequency, varying smoothly from 40 MHz to 120 MHz (Figure 8). The gain ranges from 6.2 dBi at 40 MHz to 7.2 dBi at 100 MHz then dropping back to 6.0 dB at 120 MHz.

*4.2 Receiver*

The DARE Receiver (Figure 7) provides amplification of the antenna signal to a level sufficient for further processing by the Digital Spectrometer and incorporates a load switching scheme to assist calibration. There are two receivers to accommodate the Science Antennas – one per antenna polarization. The front-end electronics will reside within the central mast of the dipole structure so that the amplifiers can be placed in close proximity to the Antenna to ensure the simplest possible interconnection. This strategy is based on EDGES' experience, with the goal of minimizing the complexity of the receiver noise model, especially with regards to spectral artifacts that would otherwise be introduced by longer cables and other reactive elements between the Antenna and amplifier.

Load switching is at the core of differential radiometry. In the DARE receiver, the switching is performed directly after the first stage low noise amplifier (LNA). Hence, the two pathways through the switch consist of 1) the antenna followed by an LNA and 2) the load resistor followed by an LNA. The two LNAs form a matched pair sporting nearly identical bandpass behavior. The DARE receiver also incorporates the novel feature where, for a given polarization, each of the two arms or cones of the bi-conical dipole is connected to a separate LNA. The signals exiting the switches are combined by a 180 degree hybrid coupler to form the output signal. Hence, for the complete dual polarization system, there are two hybrid couplers and four each of the antenna arms, load resistors, matched pairs of LNAs, and RF switches.

Although this circuit configuration appears similar to a Dicke switch for a microwave radiometer, it differs substantially from a conventional load switching radiometer to meet the science objectives. These switches serve primarily as spectral references, and secondarily for temperature calibration. The switches will be operated both in tandem and independently to confirm that spectral changes observed in each Antenna are symmetric and uncorrupted by spectral artifacts that may be introduced downstream. By placing the switch after the first LNA, the LNA itself serves as the isolator to keep switch impedance constant throughout the switching cycle. This provides an effective tool to monitor the gain stability of the system. The gain of the system is further ensured by controlling the thermal environment of the front end (discussed below).

Noise diodes further down the receiver chain provide a reliable additive noise temperature during operation. These noise diodes will be used to monitor spectral response and linearity of downstream components. The inclusion of these circuits is based on many years of radiometer experience at JPL including the radiometer for the JUNO mission and the advanced microwave radiometer (AMR). By including redundant and independent calibration sources, problems can be quickly detected and diagnosed throughout the development, testing, and operation phases. Section 4.4 discusses further details of the instrument calibration methodology.

The signal will be amplified further in balanced stages, employing 180° hybrids to split the signal into opposing phases and combining them again after amplification. This well-known topology ensures symmetry in the voltage transfer and thereby minimizes distortions otherwise present in single-ended amplifier configurations. This is a precaution against spectral artifacts that would otherwise be produced by intermodulation of the signal with itself. Again, maintaining linearity and spectral purity is paramount.

The spacecraft thermal environment is expected to vary from −100 C to +130 C. If not thermally controlled, gain variations within the Receiver would corrupt the science measurement. Critical front-end components will be packaged in a "cylinder within a cylinder" thermal system, developed and demonstrated at JPL for ground-based and space-borne RF systems, that maintains mK stability. The concept involves placing the front-end electronics within the mast that supports the Antennas. Heat fluxes from the dipole elements are shunted at the base of this assembly, which is actively stabilized with proportional-integral-differential (PID) temperature controllers operating Peltier elements to pump heat downward in two stages - into the mast structure and eventually to spacecraft radiators. The top of the mast structure contains the active (i.e., heat producing) receiver electronics, which are thermally shunted downwards into this same thermal control point. The top enclosure is constructed much as a Dewar, with 3 concentric polished metal cans forming a thermal break. The extremely tight 1 mK thermal stability is achieved in stages: the outer mast surface will vary by about 30 K between sunlight and darkness, according to preliminary analysis; the intermediate thermal shunt will vary ±0.1 K, and the inner electronics by ±1 mK. The mass within this thermal enclosure is estimated to be 0.1 kg, and electrical wiring will be thermally shunted to the heat sinks at each control point via heat conducting electrical insulators between each stage.

The remainder of the Receiver will be housed within the spacecraft. Cryogenic cables will act as a thermal break to prevent heat losses from the Receiver enclosure. This type of cable has heritage from deep space, near Earth, and military applications; has superior insertion loss and phase performance throughout a large temperature range; and has a tight bend radius, allowing for ease of assembly. A final stage of amplification will provide sufficient signal amplitude to drive the analog-to-digital conversion in the Digital Spectrometer. Low-pass filtering of the signal also will be performed to prevent aliasing within the Digital Spectrometer.

A (standard) electronics assembly will provide regulated ±12 volt DC power to the front-end assembly. It is important to regulate the voltage externally to the front end so that heat dissipation within the front end is constant with time. An onboard 3-terminal linear regulator will reduce the 12V DC to the desired bias voltage with appropriate filtering to minimize bias-related noise. The switches and noise diodes will be biased using separate linear regulators, their control circuitry will be inside the front-end, and the control lines will be activated via opto-couplers to enhance noise immunity. The temperature sensor outputs will have a dedicated connector to minimize noise contributions. The packaging will have sufficient EMI shielding to minimize external influences and prevent any surface charge related incidents.

*4.3 Digital Spectrometer*

The Digital Spectrometer design is driven by requirements to resolve mK level spectral features on smoothly varying backgrounds ranging from about 4000 K at 40 MHz to about 400 K at 120 MHz and to provide a spectral resolution of better than 10 kHz, with excellent channel-to-channel isolation. These requirements are modest by modern standards, as are the resulting data rates.

For the DARE Radiometer, a digital polyphase spectrometer has been adopted. Such spectrometers can be designed to the required bandwidth, resolution, channel shapes, and Analog-to-Digital Converter (ADC) resolution, and can be implemented using a space-qualified FPGA. Jarnot et al. (2011) describes a prototype, including field data.

There are several suitable rad-tolerant ADCs, such as those from National Semiconductor, that implement two ADCs, allowing a dual Digital Spectrometer to be implemented in a single ADC and FPGA device.

There are many FPGA-based digital polyphase spectrometers, with systems potentially able to use either a Xilinx or Actel FPGA. Both are available in a radiation-tolerant form; the Actel implementation offers potentially reduced power consumption. On-going work at JPL includes modeling existing FPGA-based digital polyphase spectrometers in order to develop a suitable design, using either a Xilinx or Actel FPGA.

FPGAs can also implement a soft-core processor, allowing some C&DH functionality to be included in the spectrometer. This allows a degree of "intelligence" and programmability, e.g., the ability to ignore portions of the spectrum subject to any internal interference and to bin channels intelligently to minimize data rate.

*4.4 Calibration*

The Instrument design is coupled to a multi-tiered calibration strategy to obtain the RF spectra from which the 21-cm signal can be extracted. The calibration strategy relies on the fact that a high degree of absolute calibration is not required, but focuses on the relative variations between spectral channels, which are much easier to control. This multi-tier strategy is based on proven techniques for receiver calibration and augmented by techniques developed during four years of design refinement and thousands of observational hours with EDGES.

*4.4.1 Calibration of the Antenna-Receiver Junction*

The relative spectral smoothness requirement presents the largest challenge at the Antenna-Receiver interface. Based on EDGES experience, DARE reduces significantly the transmission line length between the antenna and first-stage amplifiers. This arrangement causes any reflected power to create only long-period standing waves that can be fit with a smooth function in frequency.

An impedance mismatch between the antenna and first-stage amplifiers results in a spectrum,

$$T_{ant}(\nu) = [1 - |\Gamma(\nu)|^2]T_{sky}(\nu) + [1+2\varepsilon|\Gamma|\cos(\beta) + |\Gamma(\nu)|^2]T_{rcv}(\nu), \qquad (3)$$

where $T_{sky}$ is the sky brightness convolved with the antenna beam, $\Gamma$ is the reflection coefficient of the antenna due to the impedance mismatch (Harker et al. 2011). $T_{rcv}$ is the power that propagates out of the input of the first-stage amplifiers back to the antenna, $\beta$ is the phase shift due to the antenna's electrical path length and embedded circuitry, and $\varepsilon$ is the correlation coefficient between the noise emitted at the input and output ports of the first-stage LNA. The first term in this equation represents the antenna power propagated into the receiver, while the second term is due to the noise coming directly from the receiver, combined with that reflected from the impedance mismatch. The terms in the second set of square brackets are due to correlations between reflected receiver noise and noise that propagates directly from the output of first-stage amplifiers into the rest of the receiver.

For a broadband instrument such as DARE, the antenna's impedance varies strongly with frequency and, in general, will not match the receiver's impedance. However, the antenna impedance can be measured (or modeled) to sufficient accuracy to satisfy the 10% absolute calibration requirement, and the antenna's compact structure also means the impedance will change slowly with frequency.

Moreover, placing the first-stage amplifiers close to the antenna feeds results in an effective electrical path length of less than $0.1\lambda$ between the antenna and amplifiers, making the period of the spectral standing wave much larger than the instrumental bandwidth, and thereby reducing the number of degrees of freedom needed to fit the spectral structure.

*4.4.2 Calibration of Analog & Digital Path*

The Instrument must achieve a dynamic range exceeding $10^6$ in its final integrated spectrum between the average sky noise power and the desired spectral signal (1 mK signal with about a 1000 K foreground), but it only needs to calibrate the absolute power at the 10% level. This dynamic range and absolute calibration precision are regularly achieved in ground-based and space-borne receivers using internally-switched inputs to calibrate the gain and additive noise of the analog and digital receiver path. EDGES demonstrated that this calibration approach, applied in a similar spectral band, can achieve a dynamic range greater than $10^5$. The EDGES results were limited not by instrument performance, but by available time.

The Instrument follows the traditional internally-switched calibration approach to correct for additive noise and linear gain variations in the receiver path by switching the input between the antenna feeds and two loads (Figure. 7). Toggling between the loads and the first-stage amplifiers produces a three-state system,

$$p_0(\nu) = g(\nu) [p_{atten}(\nu) + p_{amp}(\nu)]$$
$$p_1(\nu) = g(\nu) [p_{load}(\nu) + p_{amp}(\nu)] \qquad (4)$$
$$p_2(\nu) = g(\nu) [p_{ant}(\nu) + p_{amp}(\nu)]$$

where $p_0$, $p_1$, and $p_2$ are the spectra from the three input states. Here, $g$ is the collective gain of the amplifiers and filters in the analog path, $p_{atten}$ is the spectrum from an attenuated load, $p_{load}$ is the spectrum from the load without attenuation, $p_{amp}$ is the spectrum from the first-stage LNA, and $p_{ant}$ is the sky spectrum after propagating through the antenna. Using the pre-flight temperature calibration of the noise loads, the three spectra are used to calibrate the antenna temperature,

$$T_{ant}(\nu) = T_{atten}(\nu) + [T_{load}(\nu) - T_{atten}(\nu)][p_2(\nu) - p_0(\nu)]/[p_1(\nu) - p_0(\nu)] \qquad (5)$$

where $T_{load}$, and $T_{atten}$ are the pre-flight calibrated noise spectra and $T_{ant}$ is the calibrated antenna temperature. Spurious signals produced by the digital electronics cancel through the subtractions in Equation 5 while the divisions remove the gain contribution $g$.

In the current design, the Instrument cycles between the 3 states on a regular cadence of approximately 60 s during science observations. The resulting spectra are transmitted to the ground data system, which implements Equation 5 to yield calibrated spectra. Equal time will be spent in each state so that the thermal uncertainty in the final calibrated spectrum is not dominated by a single state.

Pre-flight calibration of the first-stage loads is needed only at 10% to satisfy the absolute calibration requirements. During flight, control and monitoring of the load temperatures ensure that the pre-flight calibration can be applied. A compact, clumped element load was shown by EDGES to provide a suitable, spectrally smooth reference spectrum.

*4.4.3 Modulation of the Sky*

The observing strategy is a final means for precision calibration. By treating equation (5) also as a function of spacecraft orientation, $T_{sky}$ becomes dependent on four variables (frequency, roll, pitch, yaw), whereas the instrument-dependent systematic effects remain independent of the spacecraft orientation and only depend on frequency. Spacecraft orientation changes modulate the sky signal by viewing different regions of the Galactic emission. This modulation breaks the degeneracy between the instrumental terms and the sky signal (equation 5) and allows the reflection coefficient and other calibration effects to be fit as model components in a comprehensive analysis of the data time series, much like the approach employed to calibrate many CMB experiments (Section 5). Any orientation-dependent structure in the spectrum that does not correlate with this modulation is isolated in the final processing since it cannot originate from the 21-cm signal.

**5. Extracting the Cosmic 21-cm Signal in the Presence of Bright Foregrounds**

Achieving the DARE science objective will be challenging because the redshifted 21-cm signal is only one contribution to the measured RF spectra. The magnitudes of these other contributions ("foregrounds") exceed that of the 21-cm signal by factors of $10^5 - 10^6$, depending upon the frequency and position on the sky.

The DARE Science Instrument measures spectra between 40 and 120 MHz. Each spectrum is the sum of three components: the 21-cm signal, foreground radiation, and Gaussian thermal noise. At each time, the magnitude of the foregrounds depends on the direction in which the spacecraft is pointing, being an average of the true sky temperature over the antenna power pattern (Figure 6). In principle, the power pattern affects the contribution of both the 21-cm signal and the foregrounds, but in practice it is so broad, and thus averages over such a large volume of the high redshift Universe, that DARE will see the same 21-cm signal in every pointing direction. The radiometer modifies the sky spectrum through a frequency-dependent instrument response (Figure 8), changes the properties of the noise, and potentially introduces spurious spectral features. We have demonstrated via a numerical simulation an effective procedure for extracting the 21-cm signal from these spectra, including all foregrounds and the instrument response, which produces realistic errors (0.5-10%) on the inferred position of the spectral Turning Points. Table 2 summarizes our procedure, and Figure 9 presents the results. The rest of this Section describes the details behind Figure 9.

DARE's antenna power pattern covers approximately 1/8 of the sky, depending on frequency, and consequently we model the DARE data set as consisting of spectra from 8 independent regions on the sky. These spectra are used jointly in our maximum likelihood analysis. Parameterized models are used for the foregrounds, 21-cm signal, and instrumental response. These parameters are constrained simultaneously in a Bayesian framework by finding the parameter combination which maximizes the posterior probability of the model parameters given the data. We use the well-established and efficient Metropolis-Hastings algorithm to map out the likelihood. This is a member of a family of Markov Chain Monte Carlo (MCMC, Lewis & Bridle 2002) simulation methods. Given a generative model that predicts the sky brightness as a function of parameters describing the contribution from foregrounds, 21-cm signal, and instrumental response, the MCMC simulation determines the most probable values of the model parameters, and confidence intervals thereon, given DARE's observations of the sky. This method produces unbiased random samples from the posterior probability distribution, with the density of samples within a given region being proportional to the probability. More likely regions of parameter space have a greater density of points, and an $x$% confidence region is a region with a density of points above a certain threshold, such that the region contains $x$% of the points in the random sample generated by MCMC.

The choice of model parameters is crucial. We must choose a parameterization sufficiently flexible to describe all contributions to the integrated spectra accurately, yet without introducing severe degeneracies between the parameters. The 21-cm signal is parameterized by the frequency and amplitude of Turning Points B, C and D. Figure 9 illustrates the constraints on these Turning Points from a simulated DARE mission . Figure 10 shows how uncertainties on these parameters translate into uncertainties on the shape of the recovered 21-cm signal. The different parameter groups are summarized in Table 3, and we describe them in more detail in Section 5.1.  For further details on the MCMC methodology applied to the 21-cm sky-averaged signal problem, see Harker et al. (2011).

*5.1 Models for the foregrounds and the DARE frequency response*

The Turning Point parameterization of the 21-cm signal (parameter group 1 in Table 3) has been described in Section 2. Because each of our eight sky regions covers a very large volume of the Universe at the redshifts of interest to DARE, the 21-cm signal spectrum averages out and is identical in each region, so a total of 6 parameters is required to describe the 21-cm signal. By contrast, the foregrounds are spatially varying, as can be seen in Figure 11.

Most of the spatial structure apparent in Figure 11 is synchrotron radiation from the Galaxy and constitutes the dominant foreground, responsible for approximately 73% of the total (Shaver et al. 1999). Synchrotron radiation should vary smoothly in frequency (e.g., Petrovic & Oh 2010). A second contribution to the foregrounds comes from extragalactic sources. Though these are discrete, the solid angle of the DARE beam is so large that it averages together a large number of sources, so that they can be treated as another diffuse foreground. These sources also have a synchrotron spectrum. A final, small (~1%) contribution to the diffuse foregrounds comes from thermal (free-free) radiation from our Galaxy, which also should produce a smooth spectrum. Because of this smoothness, the diffuse foreground spectrum in each sky region can be described using only a small number of parameters. The model used is

$$\log T_{FG} = \log T_0 + a_1 \log \nu + a_2 (\log \nu)^2 + a_3 (\log \nu)^3 \qquad (6)$$

where $\{T_0, a_1, a_2, a_3\}$ are the parameters. Higher order polynomials are trivial to incorporate, but Pritchard & Loeb (2010) concluded that this set of parameters was the smallest number of polynomial coefficients needed to describe the spectrum to the required level of accuracy. In general, the parameter values are different in each of the 8 sky regions, making 32 diffuse foreground parameters in total (group 2 in Table 3).  This variation across the sky is an important component of our measurement approach:  It is possible to separate the foregrounds from the 21-cm signal because the foregrounds are spectrally smooth while the 21-cm signal has spectral features, and the foregrounds are spatially varying while the 21-cm signal is not (at DARE's resolution).

The Sun also makes a ~1% contribution to the foregrounds. It is small because it is diluted over the large DARE antenna power pattern. The contribution of the Sun to each pointing is different, given both its different sky positions as a function of time and possibly varying amplitude as a function of solar cycle, but the spectral shape should remain the same. The quiet Sun has a smooth thermal spectrum, which we describe using the same model as the diffuse foreground. However, $a_1$, $a_2$, and $a_3$ are the same in each sky region, while $T_0$ differs, making a total of 11 parameters (group 3 in Table 3). The Sun can also produce strong (non-thermal) radio bursts. Our approach is not to acquire science data during these relatively short intervals (between a few seconds and up to ~1 hr, Wild et al. 1963), unless the Sun is occluded by the Moon. There are approximately tens of such bursts per year (Gopalswamy et al. 2008), depending on the phase in the solar cycle, and hence they will affect only a small proportion of the mission. Their strength means that bursts can be easily identified, and those data excluded from further processing.

The thermal emission of the Moon enters the back lobe of the antenna, and is described by a single model parameter (the temperature). The Moon also reflects radiation from the sky back towards DARE, and the strength of this reflection is described by one more parameter, making a total of 2 parameters for the properties of the Moon (group 4 in Table 3).

The foreground and signal spectra are modulated by the frequency response of the instrument, giving a final antenna temperature which is well modeled by equation (3) (see Figure 8). The DARE Science Instrument is designed to make the reflection coefficient, $\Gamma$, as smooth as possible, so that the expected spectral smoothness of the foregrounds can be exploited to extract the 21-cm signal (note that the antenna reflection coefficient should not be confused with the parameter describing the reflectivity of the Moon). We model the amplitude of the reflection coefficient, $|\Gamma(\nu)|$, and the phase term, $\beta(\nu)$, as the ten lowest frequency coefficients of their discrete cosine transform. This was found to be the lowest number of coefficients which allowed us to model the smallest-scale features of $\Gamma(\nu)$ computed from an electromagnetic model of the antenna. The discrete cosine transform was chosen as a simple and fast way of modeling $\Gamma(\nu)$ as a sum of orthogonal functions. It was chosen after some experimentation with various transforms as the one which fit the computed $\Gamma(\nu)$ to reasonable accuracy using a small number of coefficients. Along with $\varepsilon$, and a parameter for the receiver temperature, there are 22 instrumental parameters to be fit (group 5 in Table 3). Our parameter space therefore has a total of 73 dimensions (6+32+11+2+22).

Figure 9 illustrates that DARE can achieve its science objectives by obtaining the amplitudes and frequencies (redshifts) of the spectral Turning Points with high confidence in the presence of these cosmic foregrounds. This figure shows the 68 and 95 per cent confidence regions on the Turning Point parameters, after marginalizing over the nuisance parameters (i.e., those of the foregrounds and instrument).

*5.2 Other Potential Contributions*

Other processes have a minor effect on the spectrum. As DARE orbits the Moon, dust particles from the lunar exosphere impact the antenna. The plasma formed by these impacts generates an antenna response (e.g., Meyer-Vernet 1985), which also constitutes a weak foreground. The resulting spectrum is estimated from models of the dust distribution around the Moon (e.g., Stubbs et al. 2010), and from the velocity, altitude, and surface area of DARE. Figure 12 shows that the dust foreground is negligible for DARE.

The plasma that composes the interplanetary medium in the lunar neighborhood radiates at frequencies well below the DARE band (e.g., plasma shot noise). Most of these plasma effects produce radio emission near the plasma frequency which is typically <100 kHz for the solar wind (e.g., Kasaba et al. 2000, Stenzel 1989). Auroral Kilometric Radiation (AKR), generated by cyclotron emission from electrons spiraling within the Earth's magnetic field, produces emissions at frequencies 50-500 kHz (e.g., Ergun et al. 1998). Thus, local plasma effects in the Moon-Earth environment will not contribute to DARE's foreground.

Jupiter is also a radio source, but its amplitude is of order $10^{-4}$ that of the Sun (e.g., Zarka 2004), so at present it is not explicitly included in our modeling.

In addition to these spectrally smooth foregrounds, carbon atoms in the Galaxy produce radio recombination lines (RRLs, Peters et al. 2010). These lines are sharp (~10 kHz wide), but spaced at known intervals of about 1 MHz. Any spectral channels containing RRLs will need to be discarded before combining the one-second spectra to make the integrated spectra. The discarded channels will constitute a very small fraction of DARE's bandwidth and so their excision will have only a minor effect on the level of noise in the final spectra.

DARE's placement in a lunar orbit allows it to access the RFI shielded zone above the lunar farside so that it does not have to contend with human-generated RFI. However, spacecraft at the Earth-Sun L2 point could potentially reflect terrestrial (or solar) emissions at a level that would vitiate DARE's intended measurements. We estimate that any such reflections are, in fact, not likely to constitute a problem. For example, reflection of RFI off the JWST sunshade is calculated to be $10^{-7}$ that of the thermal noise in Figure 12.

## 6. Constraints on the Epochs of First Star and Black Hole Formation from DARE Observations

Accurate constraints on Turning point B are the most challenging because of the brighter foregrounds (Figure 12) and diminished instrument sensitivity (Figure 8) at lower frequencies. But, Turning Point B is important because it marks the epoch of first star formation and it will be the most difficult to constrain by any other method or mission. The DARE synthetic observation modeling (Section 5) was used to estimate the uncertainties in the redshifts and frequencies of all three Turning Points shown in Table 4. For 3000 hrs of integration, we obtain a tight (~10% of the true value) measurement of the frequency and redshift of Turning Point B. The redshifts of Turning Points C (first accreting black holes) and D (beginning of Epoch of Reionization) can be measured with small uncertainties of only ±1.4% and ±0.4%, respectively, at the 95% confidence level in this integration time.

With a 200 km altitude lunar orbit, the orbital period is 127 minutes, during which the DARE spacecraft will be out of direct line-of-sight of the Earth for 45 minutes. Diffraction effects around the lunar limb reduce the useful science collection time. For the RFI to be attenuated below the expected science signal (>100 dB), the effective collection time is reduced to approximately 30 min, consistent with experience from RAE-2 and the Apollo VHF communication system. Modeling of the thermal load of the Science Instrument as it moves in and out of sunlight above the farside indicates that no more than about 5 minutes is lost due to thermal relaxation. To be conservative, we adopt a science collection time of 15 minutes per orbit. With this conservative collection time, the baseline mission lifetime is 3 years in order to obtain 3000 hours of integration, from which the level of uncertainties in Table 4 is achieved.

## 7. Summary

The Dark Ages Radio Explorer (DARE) is a space mission concept that will probe for the first time the epoch in the early Universe that begins with the Dark Ages and leads to the Cosmic Dawn when the first stars, galaxies, and black holes were born. DARE will make unique observations of the highly redshifted 21-cm signal from the hyperfine transition of neutral hydrogen arising from the intergalactic medium. The sky-averaged spectrum is theoretically expected to contain features, which we term "Turning Points", that correspond to the redshifts (epochs) when the first stars were formed, when the first accreting black holes turned on, and when the Universe began its last phase transition from neutral to ionized hydrogen (Epoch of Reionization). DARE is proposed to orbit the Moon and to take data only above the radio-quiet lunar farside, the ideal environment for such observations. The science instrument is composed of a three component radiometer that includes a pair of bi-conical, tapered, electrically-short dipole antennas, along with a receiver and a digital spectrometer. A key feature of DARE is the calibration and foreground removal approach, derived from experience with the ground-based pathfinder EDGES, using differential spectral radiometry and a Markov Chain Monte Carlo maximum likelihood analysis tool similar to the approach used for observations of the Cosmic Microwave Background.

**Acknowledgements**

The authors would like to thank the management of the NASA Ames Research Center (ARC) for their strong support and investment in the DARE concept, especially ARC Director P. Worden along with P. Klupar, B.J. Jaroux, as well as NASA Lunar Science Institute (NLSI) Director Y. Pendleton and NLSI Deputy Director G. Schmidt. We are particularly grateful to L. Webster, J. Bauman, H. Sanchez, D. Santiago, T. Soderman, and J. Baer at ARC, as well as I. O'Dwyer, A. Tanner, and R. Jarnot at JPL. We would like to acknowledge Ball Aerospace and, especially, L. Hardaway for their investment of resources and effort on the DARE concept. This concept was conceived and supported by the Lunar University Network for Astrophysics Research (LUNAR) (http://lunar.colorado.edu) , headquartered at the University of Colorado Boulder, funded by the NLSI via Cooperative Agreement NNA09DB30A.

Table 1. A comparison of current and future space missions and ground-based instruments that will probe into Cosmic Dawn and the Dark Ages.

| Telescope | Feature Probed | Redshifts | Frequency |
|---|---|---|---|
| HST/JWST | Stellar Population | ≤ 10 (HST) – 15?(JWST) | Optical/IR |
| ALMA | Molecular Gas | <10? | Millimeter/submm |
| LOFAR | Hydrogen/end of EoR | 6–11 | 120–200 MHz |
| MWA | Hydrogen/end of EoR | 6–16 | 80–240 MHz |
| PAPER | Hydrogen/end of EoR | 6–13 | 100–200 MHz |
| EDGES | Hydrogen/end of EoR | <13 | 100-200 MHz |
| WMAP/Planck | CMB | ≈1100 | 22-90 GHz |
| **DARE** | **Hydrogen/First Stars** | **11–35** | **40-120 MHz** |

Table 2. Summary of the planned DARE data analysis procedure

| |
|---|
| 1. Combine individual (one-second) spectra to produce calibrated temperature measurements of eight independent regions of the sky. |
| 2. Utilize Markov Chain Monte Carlo code to fit the data with a model describing the 21-cm global signal, the spatially dependent foregrounds in each sky region, solar system foregrounds, and the overall instrumental response. |
| 3. Recover the maximum likelihood 21-cm signal, parameterized by its Turning Points, and the errors on this signal. |

Table 3. Summary of signal model parameter groups.

| Model Parameter Groups | # |
|---|---|
| 1. Frequency & amplitude of Turning Pts. B, C, D of the 21-cm signal. | 6 |
| 2. Spectrum of diffuse foregrounds in the 8 different sky regions. | 32 |
| 3. The shape of the spectrum of the quiet Sun, and its overall normalization in each sky region | 11 |
| 4. The temperature and reflectivity of the Moon. | 2 |
| 5. The instrumental frequency response. | 22 |

Table 4. Redshifts (z) and frequencies (ν) of Turning Points in our reference model, and uncertainties in (z, ν) after 3000 hrs of integration. Uncertainties are derived from a model of the extraction process in the presence of foregrounds, thermal noise and instrumental corruption (Sec. 5). The upper and lower bounds given are at the 95% confidence level.

| Turning Point | | True Position | 3000 hrs | | | % uncertainty |
|---|---|---|---|---|---|---|
| | | | lower (upper) bound | Best-fit | upper (lower) bound | |
| B | z | 29.74 | 28.63 | 30.12 | 34.16 | 8.3 |
| | ν (MHz) | (46.20) | (47.96) | (45.65) | (40.40) | (9.2) |
| C | z | 20.75 | 20.59 | 20.85 | 21.16 | 1.4 |
| | ν (MHz) | (65.30) | (65.78) | (65.00) | (64.09) | (1.3) |
| D | z | 13.29 | 13.22 | 13.27 | 13.32 | 0.4 |
| | ν (MHz) | (99.40) | 99.86 | (99.53) | (99.20) | (0.3) |

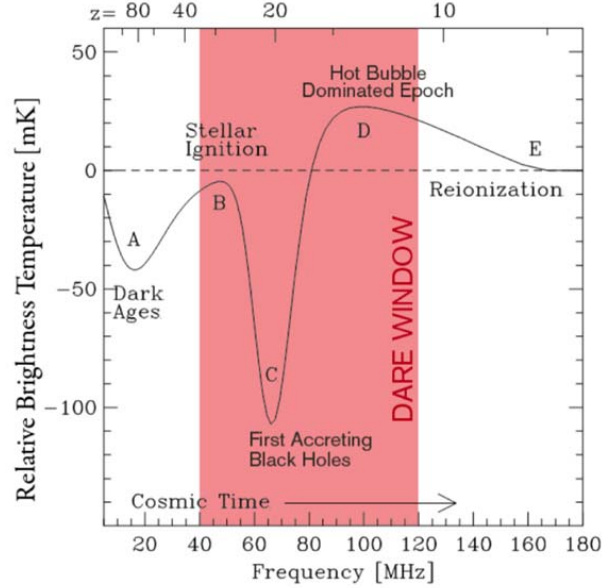

Fig. 1. Sky-averaged signal for reference model: relative brightness temperature (in units of mK, and as compared to CMB) as a function of observed frequency (bottom axis) or redshift $z$ (top axis), with key features marked. $T_B<0$ is absorption and $T_B>0$ is emission, relative to CMB. The model assumes that the properties of galaxies at $z>10$ are similar to those galaxies at $z<10$. Turning Points (A–E) mark the influence of various events on the neutral intergalactic medium, as traced by the 21-cm hydrogen line and as discussed in more depth in the text. DARE constrains the positions of the Turning Points (B, C, D) and provides the first observational constraints on the first generations of galaxies.

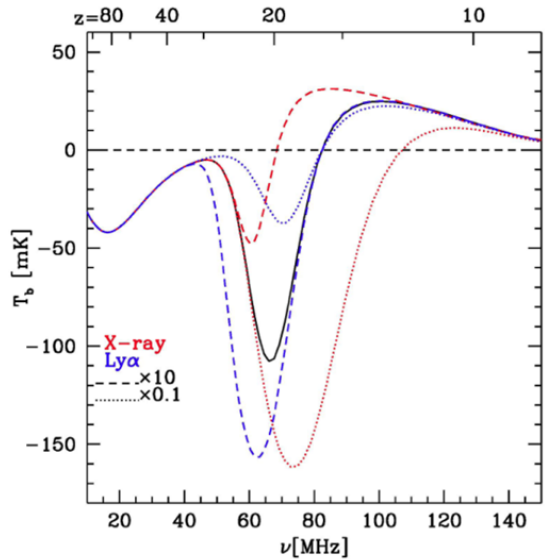

Fig. 2. Dependence of sky-averaged signal on ultraviolet flux from early stars and X-ray flux from early black holes. Solid black curve shows a reference model in which the first stars are similar to local galaxies. Blue curves show two models in which the UV flux from the first stars varies by a factor of 100, affecting the coupling of spin and gas temperature. The red curves vary the X-ray heating rate by a similar factor.

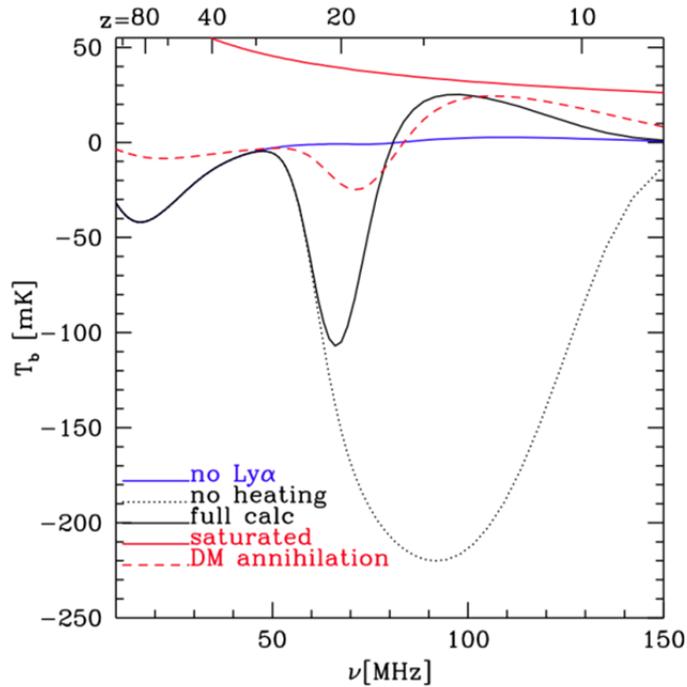

Fig. 3. Sky-averaged 21-cm signal for alternate models. Blue curve shows signal in the case that no ultraviolet (Lyman-α) photons escape from the star forming regions to couple spin and gas temperatures. Black dotted curve shows the case where X-rays are unable to heat the gas. Red solid curve shows the saturated limit of the signal. Solid black curve is our reference model. Red dashed curve shows the effect of adding extra heating from dark matter annihilation. DARE's spectrum will distinguish between these models with observations in the band 40-120 MHz.

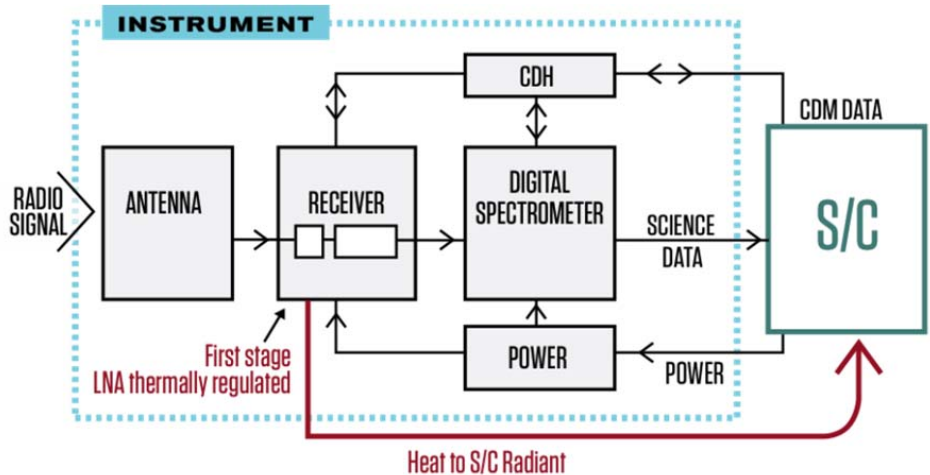

Fig. 4. Science Instrument block diagram.

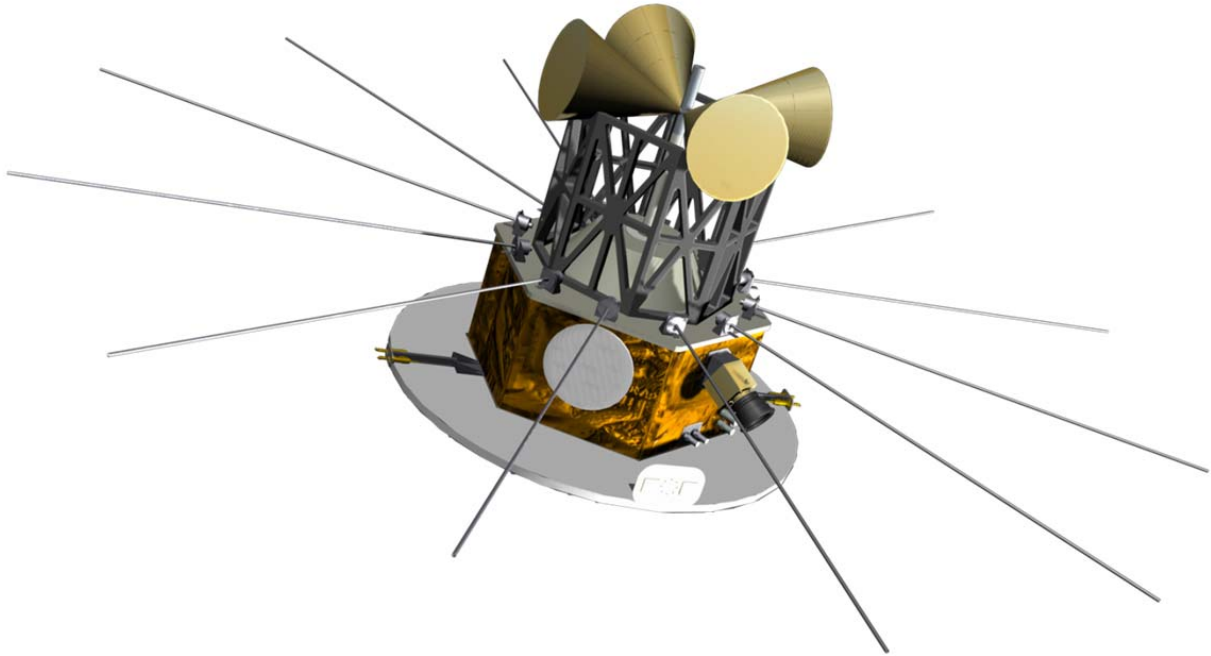

Fig. 5. Science Instrument deployed in space including bi-conical dipoles at the top, supported by a light-weight dielectric frame, the 2-m length deployed radials that serve as a ground-plane, and the solar panels at the bottom of the spacecraft.

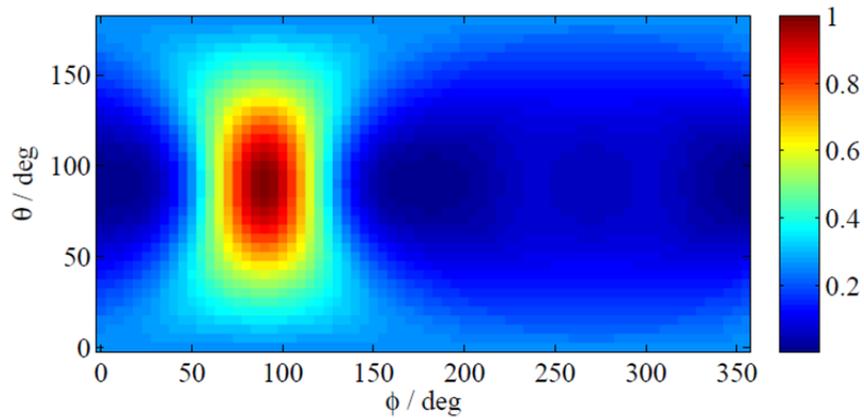

Fig. 6. Simulated antenna beam pattern as a function of angle at 75 MHz. The response is normalized to unity at the maximum, and the coordinate system is chosen such that the antenna points toward the positive y-axis ($\theta=\varphi=90°$).

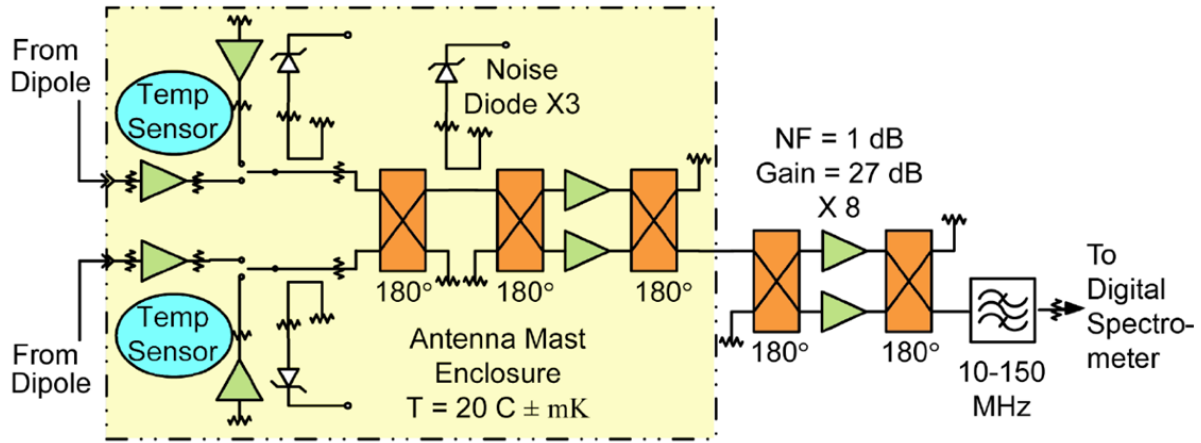

Fig. 7. DARE Receiver block diagram. Only one Receiver is shown; DARE will carry 2, one for each received polarization. All required components are available in space-qualified versions. The receiver has a noise temperature of <350K and a gain of >80dB over the 40-120MHz range. The dotted line indicates the portion of the Receiver that is controlled to ±1 mK by the "cylinder-in-a-cylinder" thermal system.

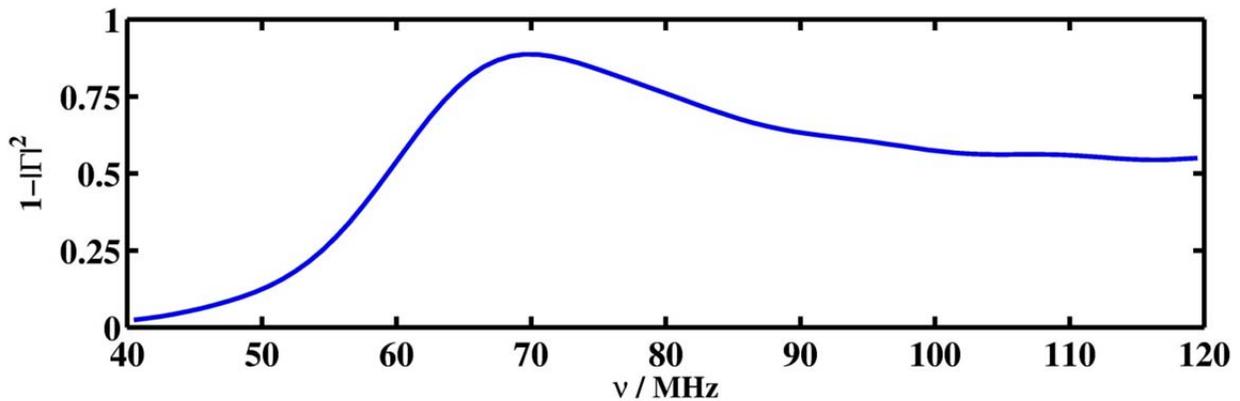

Fig. 8. Modeled DARE frequency response, $1-|\Gamma|^2$, where $\Gamma$ is the reflection coefficient (Equation 3).

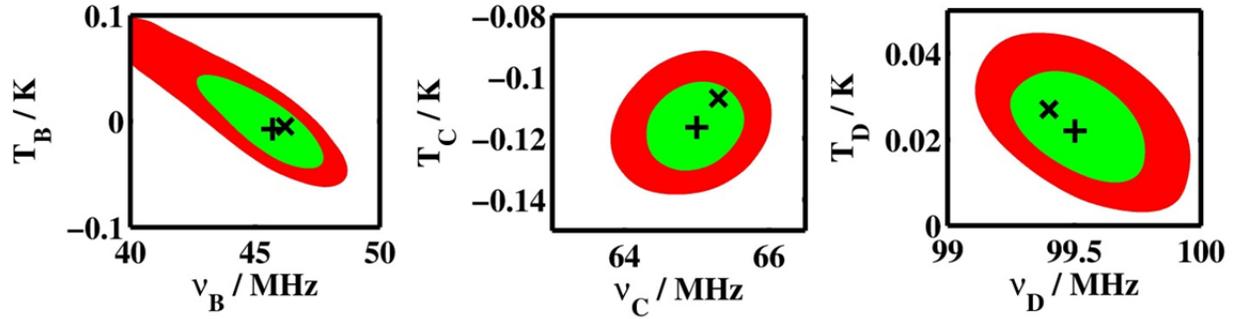

Fig. 9. Results from a DARE numerical simulation showing confidence regions on frequency and brightness temperature of Turning Points B, C & D for 3000 hrs of integration, including nominal instrument response and all cosmic foregrounds (Galaxy, discrete extragalactic sources, Sun, Moon). Green and Red regions are 68% and 95% confidence regions, respectively. An '×' marks the input model, while a '+' marks the best fit. The frequency range for Turning Point B is truncated at 40 MHz, since this is the lower limit of the DARE frequency range. Any constraints on the signal below 40 MHz are highly model-dependent, since they are obtained only from the shape of the signal at $\nu>40$ MHz. For further details, see Harker et al. (2011).

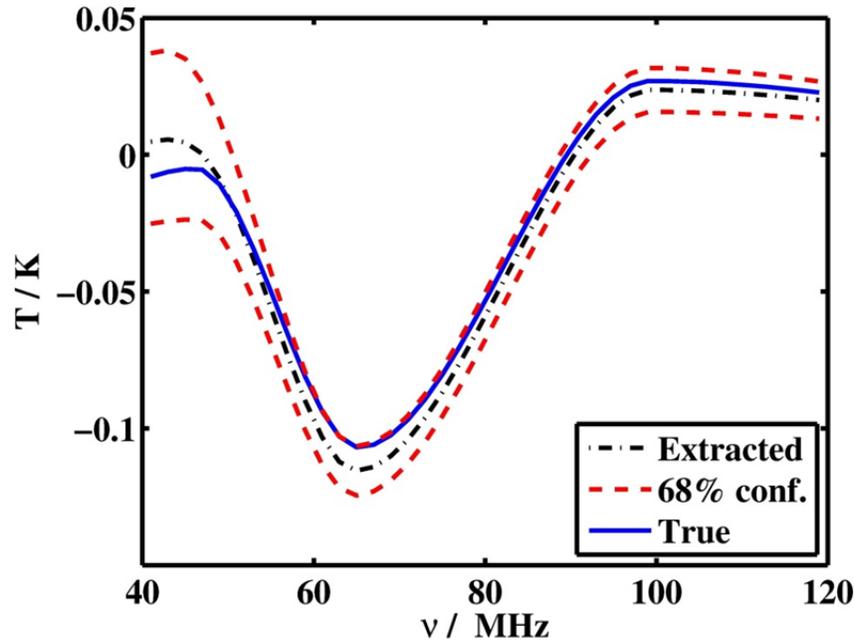

Fig. 10. Extracted signal and 68% confidence limits from a simulated baseline data set compared to the input model. The blue curve shows the input model, the black dot-dashed curve shows the extracted spectrum, and the red dashed lines show the 68% confidence regions on the extracted spectrum.

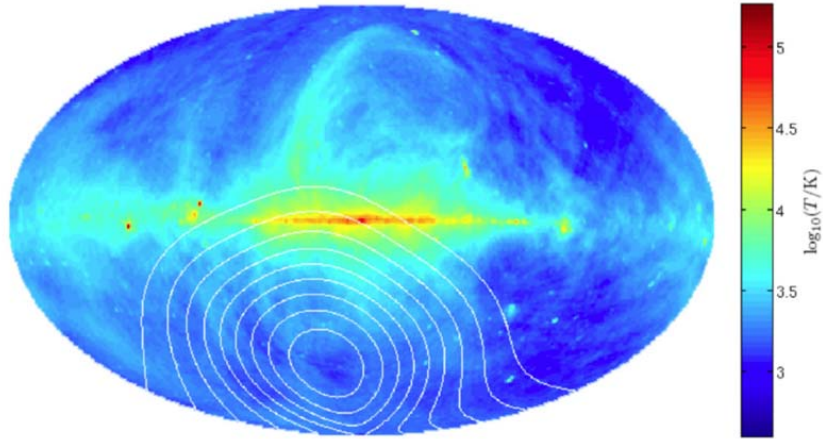

Fig. 11. Sky map at 70 MHz derived from the global sky model of de Oliveira-Costa et al. (2008). The Galactic plane is clearly visible as the intense area near the center of the map. The white contours show the power response of a crossed dipole antenna (contours spaced at 10% of maximum).

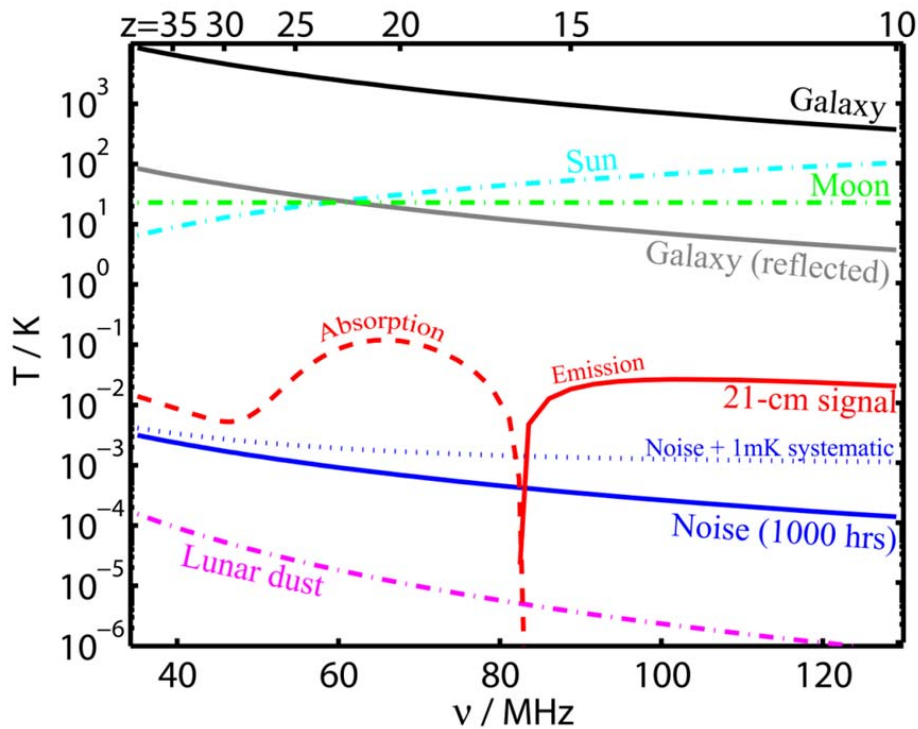

Fig. 12. Comparison of the 21-cm signal with various foregrounds. The contribution of the Galaxy is shown for a patch of sky away from the Galactic center; the reflected signal shows the magnitude of this emission after being reflected by the Moon and received through the backlobe of the antenna, which suppresses its relative contribution. Thermal radiation from the Moon is also suppressed by entering through the backlobe. The Galaxy and the Sun dwarf the 21-cm signal (shown as a red solid line when in emission, and red dashed for absorption). The blue solid curve shows thermal noise for 1000 hrs of integration, while the blue dotted curve shows this with addition of a residual of 1 mK. The noise spectrum caused by dust impacts is negligible.